\documentclass[fleqn,11pt]{article}
\usepackage{amsmath,amssymb,amsfonts,cite}

\usepackage{amsfonts,amssymb,cite}
\usepackage{graphicx}

\def\noi{\noindent}

\newcommand{\Title}[1]{\noi {{\Large\bf #1}}\\[1ex]}

\def\Aunames#1{\noi{\bf #1}}
\def\au#1{${}^{#1}$}
\def\Addresses#1{\medskip\noi \protect
	\begin{description}\itemsep -3pt {\it #1} \end{description}}
\def\adr#1#2{\item[${}^{#1}$]{\it #2}}

\newcommand{\Abstract}[1]{\vskip 2mm \begin{center}
        \parbox{16.4cm}{\small\noi #1} \end{center}\medskip}

\def\email#1#2{\footnotetext[#1]{e-mail: #2}\addtocounter{footnote}{1}}

\def\nq{\hspace*{-1em}}
\def\nqq{\hspace*{-2em}}
\def\nhq{\hspace*{-0.5em}}

\def\cm{\hspace*{1cm}}

\usepackage{color}

\def\Jl#1#2{#1 {\bf #2},\ }

\def\ApJ#1 {\Jl{Astroph. J.}{#1}}
\def\CQG#1 {\Jl{Class. Quantum Grav.}{#1}}
\def\DAN#1 {\Jl{Dokl. AN SSSR}{#1}}
\def\GC#1 {\Jl{Grav. Cosmol.}{#1}}
\def\GRG#1 {\Jl{Gen. Rel. Grav.}{#1}}
\def\IJMPD#1 {\Jl{Int. J. Mod. Phys. D}{#1}}
\def\JETF#1 {\Jl{Zh. Eksp. Teor. Fiz.}{#1}}
\def\JETP#1 {\Jl{Sov. Phys. JETP}{#1}}
\def\JHEP#1 {\Jl{JHEP}{#1}}
\def\JMP#1 {\Jl{J. Math. Phys.}{#1}}
\def\NPB#1 {\Jl{Nucl. Phys. B}{#1}}
\def\NP#1 {\Jl{Nucl. Phys.}{#1}}
\def\PLA#1 {\Jl{Phys. Lett. A}{#1}}
\def\PLB#1 {\Jl{Phys. Lett. B}{#1}}
\def\PRD#1 {\Jl{Phys. Rev. D}{#1}}
\def\PRL#1 {\Jl{Phys. Rev. Lett.}{#1}}

\def\lal{&&\nqq {}}
\def\eq{Eq.\,}
\def\eqs{Eqs.\,}
\def\beq{\begin{equation}}
\def\eeq{\end{equation}}
\def\bear{\begin{eqnarray}}
\def\bearr{\begin{eqnarray} \lal}
\def\ear{\end{eqnarray}}
\def\earn{\nonumber \end{eqnarray}}

\def\nnn{\nonumber\\ \lal }

\def\dst{\displaystyle}
\def\tst{\textstyle}
\def\fracd#1#2{{\dst\frac{#1}{#2}}}
\def\fract#1#2{{\tst\frac{#1}{#2}}}
\def\Half{{\fracd{1}{2}}}
\def\half{{\fract{1}{2}}}

\def\e{{\,\rm e}}

\def\const{{\rm const}}

\def\eqn#1{\eq\eqref{#1}}
\def\rf{\eqref}
\def\mn{_{\mu\nu}}
\def\MN{^{\mu\nu}}
\def\mN{_\mu^\nu}

\def\kappa{\varkappa}

\def\ssph{static, spherically symmetric}

\def\bh{black hole}

\def\wh{wormhole}

\def\asflat{asymptotically flat} 
\def\emag{electromagnetic}
\def\grav{gravitational}
\def\Scw{Schwarz\-schild}

\def\RN{Reiss\-ner-Nord\-str\"om}

\usepackage{color}

\begin{document}
\twocolumn[
\thispagestyle{empty}

\vspace{1cm}

\Title{On gravitating dyonic configurations in nonlinear electrodynamics}
	
\Aunames{K. A. Bronnikov,\au{a,b,c,1} S. V. Bolokhov,\au{b,2} G. S. Nurbakova,\au{d} 
		and B. Tynyshbay\au{d}}
		
\Addresses{		
       \adr a {Rostest, Ozyornaya ul. 46, Moscow 119361, Russia}
       \adr b {RUDN University, ul. Miklukho-Maklaya 6, Moscow 117198, Russia}
	\adr c {National Research Nuclear University ``MEPhI'', Kashirskoe sh. 31, Moscow 115409, Russia}
	\adr d {Al-Farabi Kazakh National University, Al-Farabi avenue, 71, Almaty 050040, Kazakhstan}
        }

\Abstract   
 {We consider static, spherically symmetric configurations of nonlinear electromagnetic fields 
   with Lagrangians $L(f)$, where $f = F\mn F\MN$, in general relativity (GR) and other metric 
   theories of gravity. The corresponding exact solutions are well known in the framework of GR 
   in cases where only an electric charge ($q_e$) or a magnetic charge ($q_m$) are present, but 
   only a few solutions in particular examples of $L(f)$ are known for dyonic systems with both 
   nonzero $q_e$ and $q_m$. We study the properties of such systems in the special case of 
   equal electric and magnetic charges and, assuming a correct Maxwell limit of $L(f)$, 
   show that there always exists such a configuration of the electromagnetic field that the 
   invariant $f$ is zero in the whole space. It leads to the existence of the corresponding 
   families of solutions both in GR in the presence of other sources of gravity (like fluids
   or scalar fields) and in a wide range of extended theories of gravity (e.g., scalar-tensor 
   and $f(R)$ gravity), in which the nonlinear electromagnetic field behaves in an 
   especially simple manner. 
}

] 
\email 1 {kb20@yandex.ru}
\email 2 {boloh@rambler.ru}

\section{Introduction}

  In general relativity (GR) and its extensions, nonlinear electrodynamics (NED) is known to be
  quite a popular source of gravity since, on one hand, strong \emag\ fields are inevitably 
  nonlinear  due to particle interactions, and, on the other, they are able to create promising 
  space-time geometries, such as nonsingular black holes and solitonlike or starlike
  configurations without horizons.

  The first examples of NED theories apparently belong to Born and Infeld whose goal was to 
  remove infinite \emag\ self-energy of a point charge \cite{B-Inf}, and to Heisenberg and Euler 
  whose theory was motivated by the physics of particle interactions \cite{EuH}. Much later, the
  NED paradigm was further developed by Plebanski \cite{Pleb} in the framework of special relativity. 
 
  The subsequent development has included a huge number of studies of self-gravitating 
  systems with NED as a source of gravity in GR and its alternatives, including searches for exact 
  solutions, their thorough studies including their stability and thermodynamic properties, particle 
  motion in nonlinear \emag\ fields in flat and curved spaces, NED causality and unitarity properties,
  and so on, see for reviews, e.g., 
  \cite{Sorokin, k-BH, Jimenez, Novello-PhRep, Gibbons, usov-11, k-dyon} 
  and references therein.

  In this paper, we focus on dyonic self-gravitating NED systems with the NED Lagrangian equal
  to $-L(f),\ f = F\mn F\MN$, possessing both electric ($q_e$) and magnetic ($q_m$) charges 
  without fixing a theory of gravity but assuming a \ssph\ space-time. 
  
  It turns out that dyonic solutions to NED-gravity equations are much harder to obtain than 
  purely electric or magnetic ones. Thus, in the framework of GR, solutions with only $q_e\ne 0$
  or only $q_m\ne 0$ are easily found with arbitrary $L(f)$ \cite{Pel-T, k-NED, FW-16, Comm-FW},
  though electric solutions are more readily obtained using a Hamiltonian-like formulation of NED, 
  the so-called P-framework \cite{Pel-T, AB-G, k-NED}, while magnetic ones do not need such a 
  re-formulation. Unlike that, for dyonic systems, solutions in GR have been obtained only for 
  specific examples of NED theories, to say nothing on extensions of GR with more complicated
  equations. 
  
  Indeed, consider $L(f)$ theory in an arbitrary \ssph\ space-time with the metric
\bearr            \label{ds} 
	ds^2 = \e^{2\gamma(u)} dt^2 - \e^{2\alpha(u)}{du^2} - r^2(u) d\Omega^2,
\nnn \cm	
			 d\Omega^2 = d\theta^2+\sin^2 \theta\, d\varphi^2,
\ear
  where $u$ is a non-specified radial coordinate and $r(u)$ the spherical radius. Then the 
  only admissible components of the Maxwell tensor $F\mn$ are $F_{tr} =- F_{rt}$ (a radial electric 
  field) and $F_{\theta\phi} = - F_{\phi\theta}$ (a radial magnetic field). The \emag\ field equations
  $\nabla_\mu (L_f F\MN) = 0$ and $\nabla_\mu {}^*F\MN = 0$ ($^* F\MN$ being the dual field) 
  then lead to
\beq              \label{F_mn}
                      r^2 L_f F^{tr} = q_e, \cm    F_{\theta\phi} = q_m\sin\theta,
\eeq
  where $L_f \equiv dL/df$, and the invariant $f$ then reads
\beq  			          \label{f-dyon}
       	f = \frac{2}{r^4(u)}\Big(q_m^2 - \frac{q_e^2}{L_f^2}\Big).
\eeq      
  If $q_e =0$, we have a simple expression for $f$ to be further substituted to the \grav\ field 
  equations. If $q_m =0$, a similar simple expression exists in the P-formulation of the theory 
  \cite{Pel-T, AB-G, k-NED}. If both charges are nonzero, then, for given $L(f)$, as noted
  in \cite{k-dyon}, \eqn{f-dyon} can be considered as either a transcendental (in general) equation
  for the function $f(r)$ or an expression of $r$ as a function of $f$. In both cases, further
  calculations are not too simple, though formally, at least in GR, a solution can be expressed
  in quadratures \cite{k-dyon}.

  In what follows, we briefly discuss some examples of the existing exact dyonic NED-GR solutions
  (Section 2) and then, in Section 3, show that with any $L(f)$ exhibiting a correct Maxwell behavior 
  at small $f$, there always exists a solution with $q_e^2 = q_m^2$ and $f\equiv 0$, in which case
  the nonlinear \emag\ field becomes effectively Maxwell, and consequently \ssph\ solutions 
  of various theories of gravity involving a Maxwell field almost automatically become solutions
  involving NED.
  
\section{Examples of dyonic NED-GR solutions}
  
  Let us begin with reproducing some known relations of GR coupled to NED. 
  In the Einstein equations with a possible cosmological constant $\Lambda$,
\beq                 \label{EE}
			G\mN - \half \delta\mN R + \delta\mN \Lambda = - T\mN
\eeq  	
  (we use units where $c = 8\pi G =1$ in standard notations), the stress-energy tensor (SET)
  $T\mN$ of the \emag\ field obeying NED with $L = L(f)$, has the form
\beq                 \label{SET}
			T\mN = -2 L_f F_{\mu\alpha} F^{\nu\alpha} + \half \delta\mN L(f).
\eeq  
  Under spherical symmetry, the SET has the only nonzero components 
\bearr    \nhq       \label{SET}
        T^t_t\! = T^u_u\! = \half L + 2E^2 L_f, \ \
        T^\theta_\theta\! = T^\varphi_\varphi\! = \half  L - 2B^2 L_f,
\nnn   
	E^2 = F_{tu}F^{ut}=\frac{q_e^2}{L_f^2 r^4}, \ \ \ 
	B^2 = F_{\theta\phi}F^{\theta\phi} = \frac {q_m^2}{r^4},
\ear
  where $E$ is the electric field strength, $B$ the magnetic induction, so that 
  $f = 2(B^2-E^2)$. Since $T^t_t = T^u_u$, the corresponding Einstein equation 
  $R^t_t = R^u_u$ is easily solved using the \Scw\ radial coordinate $u \equiv r$, with the 
  result $\alpha'+ \gamma' =0$, and under a proper choice of the time scale, denoting 
  $\e^{2\gamma} = \e^{-2\alpha} = A(r)$, the metric is written as 
\beq            \label{ds-r} 
		ds^2 = A(r) dt^2 - \frac{dr^2}{A(r)} - r^2 \,d\Omega^2.
\eeq  
  The remaining independent component $({}^t_t)$ of the Einstein equations 
  allows for finding the metric function $A(r)$ as 
\bearr          \label{A}
	A(r) = 1 -\frac{2M(r)}{r} - \frac 13 \Lambda r^2, 
\\ \lal            \label{M}
	M(r) = \frac 12 \int \! \rho(r) r^2 dr,  
\ear
  where $\rho(r) = T^t_t$ is the energy density, and $M(r)$ is called the mass function. 
  
  Now we can pass on to discussing a few particular dyonic solutions. Note that 
  in all examples to be considered, $L(f) \approx f$ at small $f$, providing a correct 
  Maxwell limit.

\medskip\noi
 {\bf Example 1:} Maxwell's electrodynamics, $L\equiv f$. Substituting $L=f$ and $L_f=1$ to 
 \eqn{M}, we obtain $2M(r) = 2m - (q_e^2+q_m^2)/r$ with an integration constant $m$, so that
\beq                    \label{RNdS}
                     A(r) = 1 - \frac{2m}{r} + \frac{q_e^2+q_m^2}{r^2} - \frac 13 \Lambda r^2,  
\eeq 
  that is, the dyonic Reissner-Nordstr\"om-de Sitter (RNdS) solution, as should be the case. The 
  electric and magnetic charges enter in the solution on equal grounds in agreement with the 
  electric-magnetic duality of Maxwell's theory.

\medskip\noi
{\bf Example 2} \cite{k-dyon}: Suppose that \eqn{f-dyon} is linear with respect to $f$, it then 
   results in a dyonic solution for the truncated Born-Infeld theory with the Lagrangian
\beq                  \label{BI}
	L(f) = b^2 \big(-1 + \sqrt{1+ 2 f/b^2}\big), \ \ \ b = \const
\eeq
  (while the full Born-Infeld Lagrangian also contains the invariant $g^2 :=(^*F\mn F\MN)^2$).

  Indeed, to make \eqn{f-dyon} linear in $f$, we must assume $L_f^{-2} = c_1 f + c_2$ with 
  $c_{1,2} = \const$. Integration of this condition gives $L = L_0 + (2/c_1)\sqrt{c_1f + c_2}$. 
  Assuming a Maxwell behavior $L \approx f$ as $f\to 0$ leads to $c_2 =1$, $L_0=-2/c_1$.
  Denoting $2/c_1 = b^2$, we finally arrive at \eqn{BI}.

  With \rf{BI}, we obtain
\bearr              \label{f-BI}
            f(r) = \frac{2b^2 (q_m^2 - q_e^2)}{4q_e^2 + b^2 r^4},
\\ \lal                   \label{E-BI}
	   \rho(r) = -\frac{b^2}{2} + \biggl(\frac{b^2}{2} + \frac{2q_e^2}{r^4}\biggr)
			\sqrt{\frac{4q_m^2 + b^2 r^4}{4q_e^2 + b^2 r^4}}.
\ear  

   The simplest case is $q_e^2 = q_m^2 = q^2$, which leads to $f=0$ and $L_f =1$, as in 
   Maxwell's theory. We then obtain $\rho = 2q^2/r^4$ and the dyonic RNdS solution with 
   $A(r)$ given by \rf{RNdS}.

   In the general case, $q_e^2 \ne q_m^2$, $A(r)$ is obtained as a long expression involving
   the Appel hypergeometric function $F_1$, we will not reproduce it here but mention some 
   of its significant features. Thus, at large $r$ the energy density $\rho$ decays as $r^{-4}$, 
   making the metric \asflat\ in the case $\Lambda =0$.
   At small $r$, by \rf{E-BI}, the invariant $f(r)$ remains finite, while 
   $\rho \approx 2 |q_e q_m|/r^4$, leading to a curvature singularity. In the pure electric 
   solution, we have again $\rho \sim r^{-4}$, while in the magnetic one $\rho  \sim r^{-2}$. 
   Thus $r=0$ is a singular center in all these dyonic solutions, in agreement with more general 
   inferences obtained in \cite{k-NED} and \cite{bok-22} on non-existence of a regular center 
   in any dyonic GR-NED solutions with $L = L(f)$.

\medskip\noi
{\bf Example 3} \cite{kru-19a}: A theory called ``Born-Infeld-type electrodynamics'':
\beq                  \label{kru1}
		L(f) = \frac{1}{\beta} \bigg[-1 + \Big(1+ \frac 43 \beta f\Big)^{3/4}\bigg], 
		\ \ \beta = \const 
\eeq
  (we have changed the sign of $L$ compared to \cite{kru-19a} to conform with our notations). 
  We thus have $L_f = (1 + 4\beta f/3)^{-1/4}$, and \eqn{f-dyon} takes the form 
\beq
		f = \frac{2}{r^4}\Big(q_m^2 - q_e^2\sqrt{1+  4\beta f/3}\Big),
\eeq   
  which reduces to a quadratic equation for $f$ that is solved to yield
\bearr                 \label{f1(r)}
		f = \frac 2{3 r^8} \big(3 q_m^2 r^4 + 4 q_e^4 \beta 
\nnn\cm
		\pm q_e^2 \sqrt {9 r^8 + 24 q_m^2 \beta r^4 + 16 q_e^4 \beta^2}\big),
\ear    
  and the expression for $\rho(r)$ to be integrated in \rf{M} has the form
\beq  \nq\,
	\rho = \frac{1}{2\beta} \bigg[\! -1 + \Big(\!1+ \frac 43 \beta f\Big)^{\!3/4}\bigg]
			+ \frac{2 q_e^2}{r^4} \Big(\!1+ \frac 43 \beta f\Big)^{\!1/4}\! ,	
\eeq  
  where $f(r)$ should be substituted from \rf{f1(r)}. Apparently, the integral in \rf{M}
  then cannot be expressed in known functions, but one point is clear, as noted in 
  \cite{kru-19a}: if $q_e^2 = q_m^2= q^2$, \eqn{f1(r)} with minus sign before 
  the square root gives $f=0$, hence $L_f =1$, and we again obtain the dyonic 
  RNdS solution with $A(r)$ given by \rf{RNdS}.  Taking the plus sign we obtain, 
  instead, 
\beq
		f = \frac{4q^2}{r^8} (3 r^4 + 4\beta q^2).
\eeq    
\medskip\noi
{\bf Example 4} \cite{kru-19b}: Arcsin electrodynamics, with
\beq   			\label{kru2}
		    L(f) = \frac 1\beta \arcsin(\beta f), \ \ \ \beta = \const,
\eeq
  with $|\beta f| < 1$. In this case, $L_f = (1\! - \! \beta^2 f^2)^{-1/2}$, and 
  \eqn{f-dyon} reads
\beq
		f = \frac{2}{r^4}\big[q_m^2 - q_e^2 (1- \beta^2 f^2)\big],
\eeq   
  which is a quadratic equation for $f$ giving
\bearr                 \label{f2(r)}
		f(r) = \frac{r^4 \pm \sqrt {r^8 + 16\beta^2 q_e^2 (q_e^2 - q_m^2)}}{4 \beta^2 q_e^2},
\ear    
  while $\rho(r)$ reads
\beq  
		\rho(r) = \frac {\arcsin(\beta f)}{2\beta} + \frac{2 q_e^2}{r^4} \sqrt{1 - \beta^2 f^2},
\eeq  
  where $f(r)$ should be taken from \rf{f2(r)}. The integral in \rf{M} again seems to be 
  not expressible in known functions. As follows from \rf{f2(r)}, the solution splits into two
  branches, depending on the sign before the square root. With the minus sign,
  the assumption $q_e^2 = q_m^2$ leads to $f=0$ and $L_f =1$, hence to
  the dyonic RNdS solution with $A(r)$ given by \rf{RNdS}. With the plus sign, the 
  solution is quite different, including the case $q_e^2 = q_m^2$ that gives 
  $f = r^4/(2\beta^2 q_e^2)$, which does not look physically interesting.  

\medskip\noi
{\bf Example 5} \cite{kru-19c}: Logarithmic electrodynamics, with 
\beq   			\label{kru3}
		    L(f) = \beta^2 \log \Big(1 + \frac{f}{\beta^2}\Big),\ \ \ \beta = \const,
\eeq
   so that $L_f = (1+ f/\beta^2)^{-1}$, and \eqn{f-dyon} is again quadratic in $f$,
\beq
		f = \frac{2}{r^4}\big[q_m^2 - q_e^2 (1+ f/\beta^2)^2\big].
\eeq      
   Its solution is
\bearr            \label{f3(r)}
		f(r) = \frac{\beta^2}{4 q_e^2} \Big(-4 q_e^2-\beta^2 r^4
\nnn \cm\cm		
		\pm \sqrt{8 q_e^2 (2 q_m^2+\beta^2 r^4)+\beta^4 r^8}\Big).
\ear
   For the density $\rho$ we have 
\beq  
		\rho(r) = \frac{\beta^2}{2} \log \Big(1 + \frac{f}{\beta^2}\Big) 
			+ \frac{2 q_e^2}{r^4} \Big(1 + \frac{f}{\beta^2}\Big),
\eeq  
  with $f(r)$ substituted from \rf{f3(r)}, and the integral in \rf{M} again looks too 
  complicated. As in the previous example, there are two branches of the solution 
  according to the ``$\pm$'' sign in \rf{f3(r)}. At $q_e^2 = q_m^2$ we obtain $f=0$
  with the plus sign and $f = -2\beta^2 (1 + \beta^2 r^4)$ with the minus sign.
  
\medskip\noi
{\bf Example 6} \cite{kru-23}: ``Rational electrodynamics,'' with
\beq   			\label{kru4}
		L(f) = \frac{f}{1 + 2\beta f}, \ \ \ \beta =\const,
\eeq
    so that $L_f = (1 + 2\beta f)^{-2}$. While this form of the Lagrangian looks quite
    simple, it happens that \eqn{f-dyon} for determining $f(r)$ is now quartic instead of 
    quadratic ones met in the previous examples:
\beq           \label{f4(r)}
		f = \frac{2}{r^4}\big[q_m^2 - q_e^2 (1+ 2\beta f)^4\big].
\eeq      
   Its solution is rather cumbersome (see \cite{kru-23}) and gives no hope to obtain 
   a complete description in known functions. It is clear, however, that $f =0$ is one of the 
   solutions to \eqn{f4(r)} in the case $q_e^2 = q_m^2$, leading as a result to the RNdS 
   configuration.

\section {A general observation for\\ NED-gravity equations}

   From the above examples (and probably others) it follows that dyonic NED-GR solutions
   are rather hard to obtain: even if $f(r)$ is known, finding the metric function $A(r)$ 
   requires integration that, in general, cannot be implemented in known functions and 
   needs using numerical methods. This concerns \ssph\ systems in GR, to say nothing on less
   symmetric systems and extended theories of gravity. However, each time it has turned 
   out that in the case of equal electric and magnetic charges there is a particularly simple 
   solution actually canceling the nonlinearity of NED and reducing the \emag\ field to its
   self-dual Maxwell configuration. Let us show that this did not happen by chance but 
   leads to quite a general observation.    

   Suppose that the NED Lagrangian $L(f)$ has a correct Maxwell limit, so that 
   $L(f) = f + o(f)$ at small $f$, and admits a Taylor expansion in a neighborhood of $f =0$, 
\beq          \label{L1}
			L(f) = f + \sum_{k=2}^{\infty} a_k f^k, \cm a_k = \const.
\eeq    
  Consider a \ssph\ configuration of the \emag\ field in a space-time with the metric \rf{ds},
  then \eqn{f-dyon} is valid, and in the case $q_e^2=q_m^2 =q^2$ it takes the form
\beq                 \label{p1}
		\Half f r^4 = q^2 p(f), \cm p(f) := 1 - L_f^{-2}.
\eeq
  Due to \rf{L1}, $p(f)$ is also a smooth function near $f=0$, and we can write 
\beq                 \label{p2}
		p(f) = \sum_{k=0}^{\infty} p_k f^k, \ \ \ \ p_k = \const. 
\eeq   
  Since $L_f = 1$ at $f=0$, we have $p(0) =0$, so it follows $p_0=0$ in \rf{p2}, and 
  the sum begins with $p_1$. Thus \eqn{p1} may be rewritten as
\beq                 \label{p3}
			f \bigg(r^4 - 2 q^2 \sum_{k=1}^\infty p_k f^{k-1} \bigg) =0.   
\eeq  
  This equation evidently has the solution $f \equiv 0$. We conclude that under the above
  conditions, there always exists this solution, with $f\equiv 0$ in the whole space, and,
  in particular, in GR it leads to the RNdS solution with the metric function \rf{RNdS}
  (where $q_e^2 + q_m^2 = 2q^2$).  
  Meanwhile, other solutions to \eqn{p3} are not excluded since there is no such a uniqueness 
  theorem, and we have really presented such solutions in our Examples 3--6.
  
  It should be stressed that this observation is not restricted to GR or to electrovacuum
  space-times. One can notice that \eqs \rf{ds}--\rf{f-dyon} are valid without fixing the 
  choice of the radial coordinate $u$, therefore, the above observation 
  is quite universal and is not restricted to systems with $T^t_t = T^u_u$ or to those 
  parametrized by the radial coordinate $r$. Also, there is no restriction to \asflat\
  configurations, as is already clear from using $\Lambda$ in the Einstein equations \rf{EE}. 
  
  In the framework of GR we can readily assert that many existing \ssph\ solutions with 
  a Maxwell \emag\ field may be re-interpreted as special solutions sourced by 
  arbitrary NED with $L(f)$ having a correct Maxwell limit. Among them are systems
  sourced by charged fluids with any equations of state (see, e.g., \cite{hobill, lemos-16}
  and references therein), though in systems with charged matter, at such a re-interpretation 
  one has to replace electric or magnetic charges and currents with combinations of 
  equal electric and magnetic charges and currents. This also concerns systems with 
  massless, massive or self-interacting scalar fields of any nature coupled to a Maxwell 
  field (see, e.g., \cite{penney-69, we-12, we-15}), including \bh\ and \wh\ solutions.   
  
  The same is true for solutions with Maxwell fields in other metric theories of gravity, 
  such as, for example, scalar-tensor and $f(R)$ theories \cite{kb-73, bruckman, stab-stt}.
  One can recall, in particular, that the most well-known scalar-tensor theories (STT), 
  belonging to the Bergmann-Wagoner-Nordtvedt class, can be formulated in two 
  conformally related space-times: the Jordan (original and preferred) conformal frame 
  and the Einstein frame where the theory looks as GR with a scalar field minimally coupled 
  with curvature but in general nonminimally coupled to matter. It can be explicitly 
  shown that, even though $L(f)$ NED theories are not conformally invariant, their solution 
  with $f\equiv 0$, with effectively Maxwell behavior of the theory, is simultaneously valid
  in both conformal frames. Furthermore, since $f(R)$ theories admit a reformulation as STT
  \cite{so-fa, fe-tsu}, the same is true for them as well.
  
  It is, however, not so evident whether or not our observation on $L(f)$ NED with $f\equiv 0$
  can be extended to solutions of STT and $f(R)$ gravity involving conformal continuations 
  \cite{cont1, cont2}, in which the whole Einstein-frame manifold is conformally related to 
  only a region of the Jordan-frame manifold, as happens, for example, in wormhole 
  solutions of GR with nonminimally coupled scalar fields \cite{kb-73, bar-vis}
  and in the exceptional \wh\ solution of the Brans-Dicke theory \cite{cont3}.
  This issue needs a further study.      

\section{Concluding remarks} 

  In conclusion, we would like to designate possible further extensions of our main result,
  the repeated emergence of solutions with $f\equiv 0$ in \ssph\ space-times inherent to
  various NED theories, independently from the underlying theory of gravity and 
  the presence of other kinds of matter non-interacting with the \emag\ field.
  
  It is quite natural to seek possible extensions to other space-time symmetries: for planar 
  and pseudospherical ones an extension looks quite straightforward, less evident it should
  be for cylindrical symmetry, and it seems rather problematic for axial symmetry.
  
  Similar points can be studied in multidimensional generalizations of NED, in particular,
  in spaces of even dimension $D = 2n$.  
  
  An interesting situation is expected with NED theories whose
  Lagrangians depend on two \emag\ invariants, $f = F\mn F\MN$ and 
  $g^2 = ({}^*F\mn F\MN)^2$. Such theories possess features quite different from those 
  of $L(f)$. Thus, in $L(f)$ theories it is impossible to obtain dyonic \bh\ solutions
  with a regular center, while in $L(f,g)$ theories such examples are known 
  \cite{tsuda, bok-26}. Not less important is the existence of duality-invariant theories
  in this class, allowing for obtaining dyonic solutions with an arbitrary combination 
  of $q_e$ and $q_m$ by duality rotations, starting from an electric or magnetic solution. 
  This property is well known for the Born-Infeld theory \cite{B-Inf} (see also \cite{gib-95} and 
  references therein) with the Lagrangian 
\[  \nhq
  		L(f, g) = b^2 (-1 + \sqrt{1\! + 2f/b^2\!  -  2g^2/b^4},\ \  b= \const
\]  
  (\eqn{BI} is its truncated version); it also takes place in the recently proposed ModMax 
  (Modified Maxwell) theory \cite{modm1, modm2} with the Lagrangian 
\[  
  		L(f,g) = f \cosh \gamma\! - \sqrt{f^2\! + g^2} \sinh \gamma, \ \ \gamma = \const,
\]   
  which, in addition, possesses conformal invariance.  It looks quite intriguing how 
  and under which conditions there would emerge solutions with $f=0$ where, 
  in general, the second invariant $g^2$ must not vanish.
  
  Thus there remain a number of questions of interest connected with possible simple 
  solutions in quite sophisticated theories.


\end{document}